\documentclass[pre,twocolumn,floatfix,tightenlines,nofootinbib,superscriptaddress,showpacs]{revtex4}%

\usepackage{amsmath}
\usepackage[pdftex]{graphicx}
\graphicspath{{figs/}}
\usepackage{subfigure}

\newcommand{\mathnotation}[2]{\newcommand{#1}{\ensuremath{#2}}}

\mathnotation{\vb}{\mathbf{v}}
\mathnotation{\htop}{h} 
\mathnotation{\e}{\mathrm{e}}
\mathnotation{\ldef}{\mathrel{\raisebox{.069ex}{:}\!\!=}}
\mathnotation{\rdef}{\mathrel{=\!\!\raisebox{.069ex}{:}}}
\mathnotation{\wb}{w_{\mathrm{B}}} 
\mathnotation{\map}{\mathcal{M}}
\def\d{{\mathrm d}}
\mathnotation{\tnot}{t_0}
\mathnotation{\xf}{x_{\mathrm{f}}}
\mathnotation{\xe}{x_{\mathrm{e}}}
\mathnotation{\T}{T} 
\mathnotation{\C}{C} 
\mathnotation{\Cinf}{C_\infty} 

\mathnotation{\BM}{g} 
\mathnotation{\A}{a}

\mathnotation{\visc}{\nu}
\mathnotation{\wu}{\mathcal{W}_{\mathrm{u}}}
\mathnotation{\ws}{\mathcal{W}_{\mathrm{s}}}
\mathnotation{\diff}{\kappa}
\mathnotation{\F}{f}
\mathnotation{\U}{U}
\mathnotation{\X}{x}
\mathnotation{\Y}{y}
\mathnotation{\xb}{\mathbf{x}}
\mathnotation{\ttau}{\tau}
\mathnotation{\R}{r}
\mathnotation{\N}{N}
\mathnotation{\D}{d}
\mathnotation{\la}{\lambda}
\mathnotation{\Cself}{\tilde{C}}
\mathnotation{\Le}{L}
\mathnotation{\Ri}{R}
\mathnotation{\Ds}{Ds}
\mathnotation{\h}{h}

\begin{document}

\title{Open-flow mixing: Experimental evidence for strange eigenmodes}
\author{E. Gouillart}
\affiliation{Surface du Verre et Interfaces, UMR 125 CNRS/Saint-Gobain, 93303
Aubervilliers, France}
\affiliation{Service de Physique de l'Etat Condens\'e, DSM, CEA Saclay,
URA2464, 91191 Gif-sur-Yvette Cedex, France}
\author{O. Dauchot}
\affiliation{Service de Physique de l'Etat Condens\'e, DSM, CEA Saclay,
URA2464, 91191 Gif-sur-Yvette Cedex, France}
\author{J.-L. Thiffeault}
\affiliation{Department of Mathematics, University of Wisconsin --
  Madison, WI 53706, USA}
\author{S. Roux}
\affiliation{LMT-Cachan, UMR CNRS 8535/ENS-Cachan/Univ. Paris
VI/PRES UniverSud, 94 235 Cachan, France}

\date{\today}


\begin{abstract} 

  We investigate experimentally the mixing dynamics of a blob of dye in a
channel flow with a finite stirring region undergoing chaotic advection.
We study the homogenization of dye in two variants of an eggbeater
stirring protocol that differ in the extent of their mixing region. In
the first case, the mixing region is separated from the side walls of the
channel, while in the second it extends to the walls. For the first case,
we observe the onset of a permanent concentration pattern that repeats
over time with decaying intensity. A quantitative analysis of the
concentration field of dye confirms the convergence to a self-similar
pattern, akin to the \emph{strange eigenmodes} previously observed in
closed flows.  We model this phenomenon using an idealized map, where an
analysis of the mixing dynamics explains the convergence to an eigenmode.
In contrast, for the second case the presence of no-slip walls and
separation points on the frontier of the mixing region leads to
non-self-similar mixing dynamics.

\end{abstract}
\vspace{-0.5cm}

\maketitle


\section{Introduction}
\label{sec:intro}

Many industrial situations require the mixing of very viscous fluids,
and involve mechanical stirring in order to achieve sufficient
homogeneity. We investigate here the mixing scenario of a
low-diffusivity dye in a free-surface open flow that crosses a
localized stirring region. Relevant applications concern continuous
throughflow industrial processes, for example the food processing,
glassmaking, or pulp and paper industries.

Despite its ubiquity, open-flow mixing has received comparatively much
less attention than its closed-flow counterpart in the applied
mathematics and physics literature. Efficient stirring flows promote
\emph{chaotic advection}~\cite{Aref1984, Ottino1989} in viscous fluids,
that is chaotic Lagrangian trajectories of point-like particles. The
temporal dynamics of the concentration field of a diffusive dye mixed by
chaotic advection have been addressed in various experimental and
numerical studies of closed flows~\cite{Antonsen1996, Rothstein1999,
Jullien2000, Villermaux2003, Gouillart2007, Haynes2005, Pikovsky2003,
Sukhatme2002, Fereday2002, Fereday2004}. Time-persistent patterns and
exponential decay rates of fluctuations observed in experiments and
simulations~\cite{Rothstein1999, Jullien2000, Sukhatme2002,
Thiffeault2003, Fereday2004} have been associated with an eigenmode of
the advection-diffusion operator dubbed \emph{strange
eigenmode}~\cite{Pierrehumbert1994, Liu2004, Thiffeault2004b, Haynes2005,
Gilbert2006}.  Non-hyperbolic structures of the phase space may
nevertheless affect mixing dynamics~\cite{Popovych2007}. In particular,
we have shown in recent experimental work~\cite{Gouillart2007,
Gouillart2008a} that fixed solid walls with no-slip boundary conditions
can support parabolic separation points leading to algebraic mixing
dynamics in closed flows~\cite{Chertkov2003a,Lebedev2004,Burghelea2004,
Salman2007}.

A wide class of open-flow mixing devices successfully used in
microfluidics rely on the repetition of similar passive or active
mixing elements --- such as grooved patterns~\cite{Strook2002},
serpentine channels~\cite{Castelain2001, Lin2007}, or elementary
micro-pumps~\cite{Okkels2004} --- that induce repeated stretching and
folding events as fluid particles flow through successive
mixing elements~\cite{Strook2002, Wiggins2004,
  Villermaux2008}. However, many large-scale processes of heavy
industry cannot afford the pressure drop induced by repeated geometry
changes, which is acceptable only for microscales.

\begin{figure}
\centerline{\includegraphics[width=0.99\columnwidth]{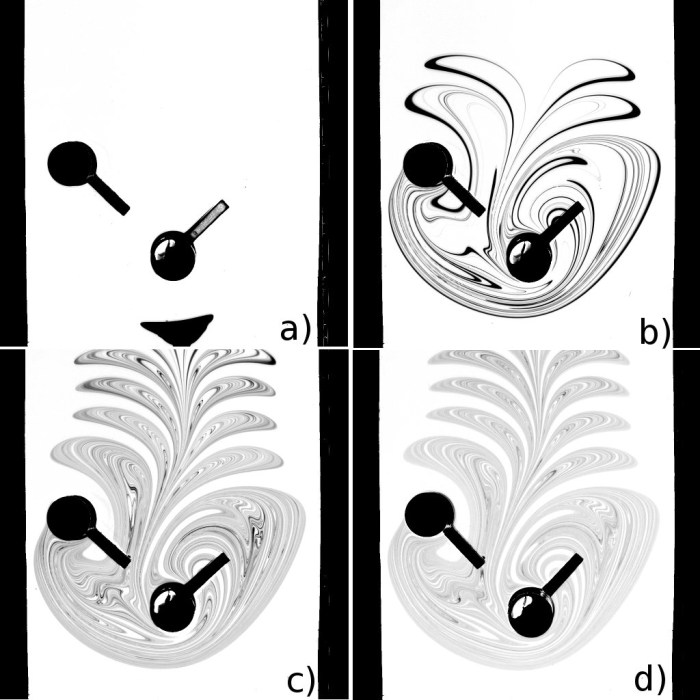}}
\caption{Stroboscoped pictures of the dye pattern (separated by integer
multiples of the stirring period),
illustrating the successive phases of a typical mixing experiment. (a) A blob of
dye is advected inside a mixing region where two rods stir passing fluid.
(b) Stirrers catch a part of the blob, which is stretched and folded into
thin filaments, while a fraction of the blob escapes quickly downstream
without experiencing much mixing. (c) and (d) After a few stirring
periods, filaments have been stretched enough to reach the diffusive
Batchelor scale, and intermediate gray levels appear in the pattern. From
this moment on, a permanent pattern appears to set in: patterns in (c) and
(d) resemble each other very closely, but with a weaker intensity at
longer times.
\label{fig:open_steps}}
\end{figure}

In contrast, we study here a free-surface channel flow that crosses a
single mixing region where stirrers fold and stretch passing fluid
particles, which is a more realistic device for large-scale
processes. In the following, by ``open-flow mixers'' we refer to
devices with a single mixing region.  In such systems chaotic
advection necessarily affects fluid particles for only a finite time,
corresponding to their stay in the vicinity of the stirring elements.
Successive stages of the mixing of an impurity --- a blob of dye ---
are shown in Fig.~\ref{fig:open_steps}: incoming dye escapes
progressively downstream, and fluid particles experience only a finite
number of stretching and folding steps. As chaos is usually
characterized by asymptotic indices such as Lyapunov
exponents~\cite{Ottino1989}, the description of finite-time chaotic
advection in open flows requires additional concepts.

Some tools of dynamical systems are nevertheless relevant to
characterize mixing in open flows. Of particular importance is the set
of periodic orbits that never escape the mixing region, called
\emph{chaotic
  saddle}~\cite{Jung1993,Pentek1995,Sommerer1996,Pentek1999,Tel2000}. This
set, together with its stable and unstable manifolds, are the backbone
of fluid transport in and out of the mixing region~\cite{Ott1993,
  Tel2000}.  During their stay inside the chaotic region, Lagrangian
particles successively shadow the trajectories of periodic
orbits~\cite{Bowen1975, Auerbach1987, Ghosh1998}, and finally leave
the chaotic region close to the unstable manifolds of such orbits. In
particular, fluid particles with long residence time in the mixing
region (Fig.~\ref{fig:open_steps}(c-d)) trace out the unstable
manifold~$\wu$ of the chaotic saddle~\cite{Tel2000}. Aspects of mixing
in open flows such as residence-time distributions, fractal dimensions
of the mixing pattern, or the dynamics of chemical reactions occurring
in the flow have been related to properties of the chaotic
saddle~\cite{Jung1993,Pentek1995,Sommerer1996,Pentek1999,Tel2000,
  Tel2005}.

However, except for an early pioneering study by Sommerer et
al.~\cite{Sommerer1996}, the evolution of the statistical properties of
the dye concentration field have never been studied quantitatively in
open chaotic flows. This contrasts with closed flows, where much research
effort has been invested in the study of persistent concentration
patterns and strange eigenmodes~\cite{Pierrehumbert1994, Liu2004,
Thiffeault2004b, Haynes2005, Gilbert2006}.  In addition, much of the
earlier work on chaotic mixing in open flows has focused on flows more
relevant for geophysics than industry (see~\cite{Tel2005} for a review),
such as Von K\'arm\'an alleys in the wake of a
cylinder~\cite{Jung1993,Sommerer1996,Pentek1999,Tel2000}. A noteworthy
exception concerns mobile point vortices in an infinite domain studied by
Neufeld and T\'el~ \cite{Neufeld1998}, and Shariff et
al.~\cite{Shariff2006}, which share similarities with moving-rod devices
that we study here.  Our paper follows similar lines to the closed-flow
studies of dye concentration cited above.  Rather than characterizing
kinematic properties of chaotic advection, we concentrate on describing
quantitatively the evolution of the concentration field of a dye tracer
over successive stirring periods in an open-flow system, shown in
Fig.~\ref{fig:open_steps}. The question of mixing efficiency and its
characterization is beyond the scope of this paper and left for future
communications.  Rather, we investigate thoroughly the evolution of the
concentration field resulting from the initial injection of a tracer.

The paper is organized as follows.  In Section~\ref{sec:exp} we
introduce our experimental apparatus, designed to reproduce some key
ingredients of open-flow industrial mixers. In
Section~\ref{sec:mix_patterns} we consider the main qualitative
features of dye patterns. In Section~\ref{sec:decay_BF}, we take on
the evolution of the concentration field for a first stirring
protocol, where mixing takes place far from the channel side
walls. After a few stirring periods, we observe the onset of a
persistent concentration pattern whose moments decay exponentially
with time, which we attribute to a concentration eigenmode.  In
Section~\ref{sec:BM}, we introduce a simplified 1-D model derived from
the baker's map~\cite{Farmer1983} to describe chaotic advection in
open flows. The onset of the concentration eigenmode and its link with
the chaotic saddle are investigated in Section~\ref{sec:symbol_dyn}
for this ideal model. In Section~\ref{sec:BS} we report on departure
from the eigenmode behavior of the concentration field in experiments
with non-hyperbolic chaotic saddles, a case that occurs for example
when the chaotic saddle extends to the fixed no-slip walls of the
channel.  We conclude in Section~\ref{sec:concl} with some final
remarks.

\section{Experimental set-up and methods}
\label{sec:exp}

We have designed an experimental device to study mixing in open flows
with two main criteria: (i) a stirring flow that promotes chaotic
advection and is simple enough to draw generic conclusions; and (ii) a
device that is realistic enough to be relevant to applications.  To
satisfy the latter criterion, we require the device to have good
mixing properties, but still
be realizable with standard mechanical components.

Viscous cane sugar syrup flows in a long glass channel of length
$2\,\text{m}$ ($\X$ direction) and width $36\,\text{cm}$ ($\Y$
direction). Sugar syrup fills the channel up to a height $\h =
5\,\text{cm}$ ($z$ direction). We fix the flow rate at
$1\,\text{L}\cdot\text{min}^{-1}$ by imposing a constant pressure drop at
the inflow of the channel, which corresponds to a velocity averaged over
a channel cross-section $\U\sim1\,\text{mm}\cdot\text{s}^{-1}$. In the
absence of stirrers, the flow is steady, and independent of $\X$
after a few viscous diffusion lengths $\U\h^2/\nu\sim 1\,\text{cm}$ from
the channel entrance. Analytical calculations and numerical
simulations~\cite{Pigeonneau_private} show that this shallow and
free-surface flow has a much flatter profile at all heights than a
Poiseuille flow, and is instead closer to plug flow save close to the
side walls. 

\begin{figure}
\begin{minipage}{0.48\columnwidth}
\centerline{\includegraphics[width=0.95\textwidth]{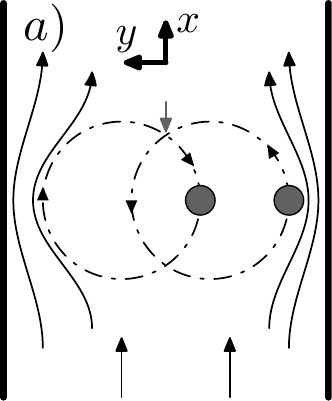}} 
\centerline{\includegraphics[width=0.99\textwidth]{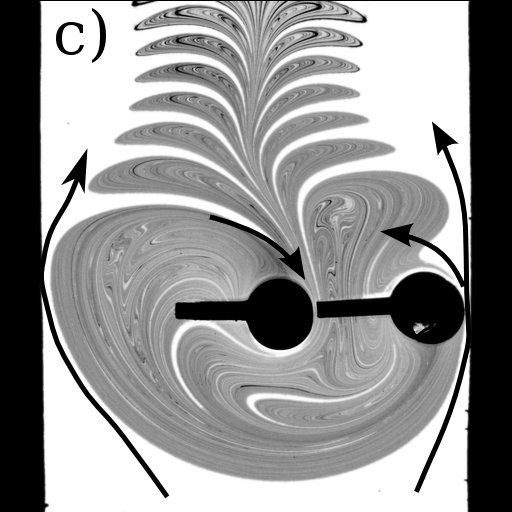}}
\end{minipage}
\begin{minipage}{0.48\columnwidth}
\centerline{\includegraphics[width=0.95\textwidth]{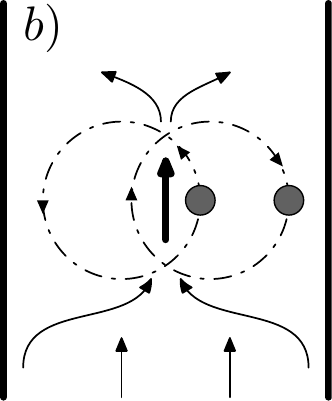}}
\centerline{\includegraphics[width=0.99\textwidth]{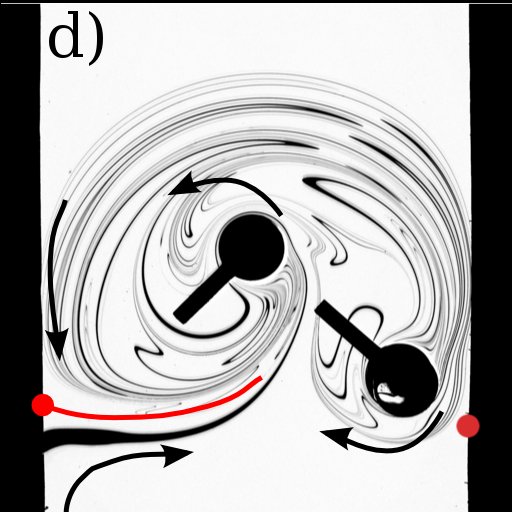}}
\end{minipage}
\caption{The butterfly ((a) and (c)) and breaststroke ((b) and (d))
  versions of the eggbeater protocol impose very different stirring
  flows, hence different mixing patterns. (a), (c) Butterfly protocol:
  the rods push fluid along the side walls. This results in a layer of
  free trajectories along each wall that are never caught inside the
  mixing region. Consequently, after a few periods the downstream dye
  pattern is well separated from the walls. (b), (d) Breaststroke
  protocol: the competition between the upstream flow and the flow
  created by the rods results in two separation points on the
  walls. This time, the mixing pattern extends over the whole channel
  width up to the walls.\label{fig:BFBS}}
\end{figure}

Our stirring protocol resembles an egg-beater device
(Fig.~\ref{fig:BFBS}). Two cylindrical stirrers of diameter
$45\,\text{mm}$ are plunged into the fluid and rotated at constant linear
speed by two motors on intersecting circular paths of diameter
$\D=140\,\text{mm}$. The distance between the centers of the two
trajectories is $105\,\text{mm}$, allowing a significant overlap of the
two circles. We can vary the stirring frequency $\F$ to tune the mean
number of stirring periods spent by a particle inside the mixing region
$\N= \D/\U\T$, where $\T=1/\F$. $\N$ is a rough measure of the typical
non-dimensional timescale of chaotic advection for Lagrangian particles.
$\N$ was varied between $4$ and $24$ to study the dependence on stirring
frequency of the evolution of the concentration field.

The direction of rotation of the stirrers with respect to the flow
direction is an important factor.  We can rotate the rods so that they
assist fluid in passing along the channel walls (Figs.~\ref{fig:BFBS}(a)
and (c)), where they tend to incorporate fluid inside the mixing region
while on the upper and central part of their trajectory (see arrows in
Figs.~\ref{fig:BFBS}(a) and (c)). By analogy with swimming, we call this
protocol the \emph{butterfly} (BF) protocol. Or we can rotate the rods in
the opposite direction, so that fluid cannot cross the stirring region
along the side walls, but is forced into a central funnel
(Figs.~\ref{fig:BFBS}(b) and (d)). To pursue the swimming analogy, we
call this protocol the \emph{breaststroke} (BS) protocol.

We inject with a syringe a spot of low-diffusivity dye (Indian ink
diluted in sugar syrup) directly beneath the surface, far
upstream of the stirring rods.  With the stirring frequencies
used here, we did not observe any 3-D motion of the dye, which
remains just beneath the fluid's surface for the duration of the
experiments.

The stirring region is visualized from below using a $2000\times2000$
\textsc{Redlake} camera of 12-bit depth. We visualize the whole width
of the channel except for a strip on each side of width
$3\,\text{cm}$ that are hidden by the frame bearing the channel.  The
region of interest (see Fig.~\ref{fig:open_steps}) is lit from above
in order to directly relate light intensity as measured by the camera
to the absorption of light by dye, hence to dye concentration. We
choose the acquisition frequency of the camera to be a multiple of the
stirring frequency. A careful calibration of the response of the
camera to dye absorption allows us to deduce from pixel values the
concentration field of dye in the whole image~\cite{Gouillart_thesis}.

\section{Mixing patterns and their link with the chaotic saddle}
\label{sec:mix_patterns}

In this section, we describe briefly some qualitative features of
mixing patterns in our experiments. In the Stokes (viscous) flow
regime, the presence of the side walls sharply dampens the stirring
flow.  We may thus consider the flow as steady and $\X$-independent
outside a \emph{mixing region}.  The typical size of the mixing region
corresponds to the range of the trajectories of the rods, plus a
damping length that scales with the channel width.  We shall now focus
on this flow region.

We obtain completely different mixing patterns by simply changing the
direction of rotation of the rods, as can be seen in
Fig.~\ref{fig:BFBS}.  However, some features of the time-evolution of
the concentration pattern --- illustrated in Fig.~\ref{fig:open_steps}
for a butterfly protocol --- are common to butterfly and breaststroke
protocols.  When the blob of dye arrives in the mixing region, some
parts of the blob escape downstream very rapidly (thick black
filaments in the upper part of Fig.~\ref{fig:open_steps}(a)). They
keep the same concentration level as the original concentration of the
blob. Thus, our mixer fails to mix such elements efficiently. However,
dye particles remaining in the mixing region (filaments caught by the
rods in Fig.~\ref{fig:open_steps}(a)) are stretched and folded by the
rods. The dynamics of Lagrangian particles are thus chaotic in at
least some part of the mixing region, leading to an exponential
separation of neighboring particles at a typical rate
$\la$~\cite{Ottino1989}. After several stretching and folding events,
the width of a filament of dye stretched at a rate $\lambda$ inside
the chaotic mixing region stabilizes at the so-called Batchelor length
\begin{equation}
  \wb\ldef\sqrt{\diff/\la}\,,
\end{equation}
where the effects of compression and diffusion balance, with $\kappa$
the diffusivity.  Obviously, stretching is not constant in the mixing
region, but the width of a filament depends only on its recent history of
stretching rate~\cite{Batchelor1959,Villermaux2003}. After
a few periods, the dye pattern therefore consists of gray dye
filaments that have reached the Batchelor scale, and of white gaps
that result from the repeated incorporation of undyed upstream fluid
(Fig.~\ref{fig:open_steps}(c)).  Most strikingly, this pattern appears
to repeat periodically in time (Figs.~\ref{fig:open_steps}(c) and
(d)), and also in space up to distortions due to the deviation from
plug flow in the channel.

The persistence of the long-time pattern visible in
Fig.~\ref{fig:open_steps} evokes permanent patterns observed in closed
flows~\cite{Rothstein1999, Jullien2000, Voth2003} that have been dubbed
\emph{strange eigenmodes}. For long times, the concentration field
converges to the strange eigenmode, which is the second slowest decaying
eigenmode of the advection-diffusion operator (the first trivial mode
corresponds to a nondecaying uniform concentration in closed flows). This
formalism was first presented by Pierrehumbert in a 1994
paper~\cite{Pierrehumbert1994}, and deeper mathematical foundations were
given by subsequent work~\cite{Sukhatme2002, Liu2004}. The strange
eigenmode regime corresponds to a \emph{global} decay of concentration
fluctuations~\cite{Haynes2005, Fereday2004}, that arises when a box of
width $\wb$ holds a large number of filaments coming from different parts
of the fluid domain.
 
We are not aware of any previous observation of strange eigenmodes in
open flows.  In the remainder of this paper, we shall examine whether the
presence of a permanent pattern, as suggested by the above visual
inspection, is confirmed by a quantitative analysis of the evolution of
the concentration field. We shall see that butterfly protocols indeed
give rise to permanent patterns, whereas the spatial distribution of dye
keeps evolving in the breaststroke case. We shall also explain the origin
of such a difference.

Important differences between mixing dynamics in the two protocols may
already be inferred from direct visual observation of
Fig.~\ref{fig:BFBS}. In the case of the butterfly protocol, after a few
periods the mixing pattern has a smooth boundary on its upstream side
and a cusped boundary on its downstream side. In addition, lobes detach
at each period from the mixing region, as the pattern is advected
downstream by the main flow and sliced by the rods
(Fig.~\ref{fig:open_steps}). Note that the lobes do not occupy the whole
width of the channel: there exists a layer of free trajectories on both
sides of the channel that are never caught by the rods and ``shield''
the mixing region from the low-stretching regions at the wall.

In the case of the breaststroke protocol, filaments are smoothly
elongated in the upper part of the mixing region, and the advected
pattern consists of roughly parallel striations of dye and white
filaments that extend over the whole width of the channel.  One of the
striking features of the breaststroke pattern is the presence of two
separation points on the channel walls (represented as small disks in
Fig.~\ref{fig:BFBS} (d)) that stem from the competition between the main
flow and the rods. Such separations points play a special role, as dye
filaments emanating from the vicinity of the separation points bear a
much higher concentration value than in the remainder of the pattern
(Fig.~\ref{fig:BFBS}(d)).

Fixed separation points observed in the breaststroke protocol are an
obvious, but atypical, example of points belonging to the chaotic saddle
\cite{Tel1990}, which is defined as the set of periodic orbits of the
flow that return to their initial position after a finite number of
stirring periods $\T$. Periodic orbits are therefore trapped in the
vicinity of the rods, unlike most trajectories.  Since Lagrangian
trajectories obey Hamiltonian dynamics in 2-D incompressible flows,
orbits of the chaotic saddle can be either hyperbolic, elliptic, or
parabolic orbits. Fluid particles are stretched exponentially with time
in the vicinity of hyperbolic points, while they loop around elliptic
points and experience at most algebraic stretching --- elliptic points
are therefore found in non-chaotic regions, also known as elliptic
islands. It can be shown~\cite{Gouillart2007, Gouillart2008a} that
separation on fixed no-slip walls are marginally unstable parabolic
points: fluid escapes their vicinity with slow, algebraic dynamics. Note
than the chaotic saddle provides an elegant way to define formally the
mixing region as the smallest connected set that contains all the
periodic orbits of the saddle.


\section{Butterfly protocol: onset of a strange
eigenmode}
\label{sec:decay_BF}

In this section, we consider the evolution of the statistical properties
of the concentration field in the butterfly protocol, and we show that
the visual observation of a concentration eigenmode is confirmed quantitatively.

\begin{figure}
\begin{minipage}{0.38\columnwidth}
\subfigure[]{
\centerline{\includegraphics[width=0.95\textwidth]{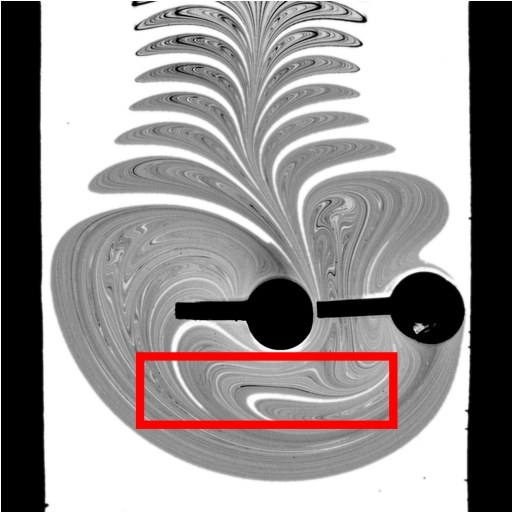}}
}
\end{minipage}
\begin{minipage}{0.58\columnwidth}
\subfigure[]{
\centerline{\includegraphics[width=\textwidth]{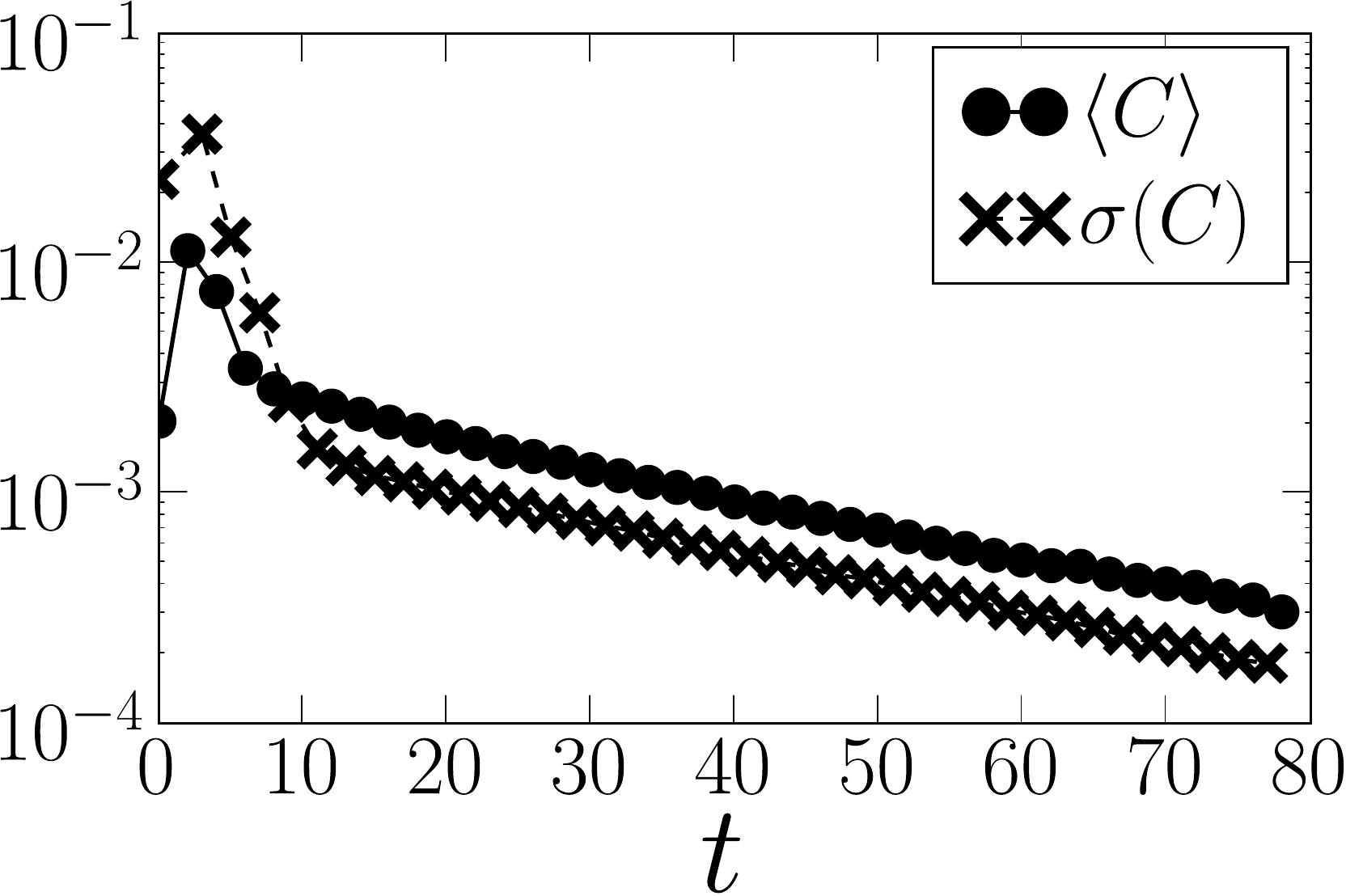}}
}
\end{minipage}
\begin{minipage}{0.58\columnwidth}
\subfigure[]{
\centerline{\includegraphics[width=\textwidth]{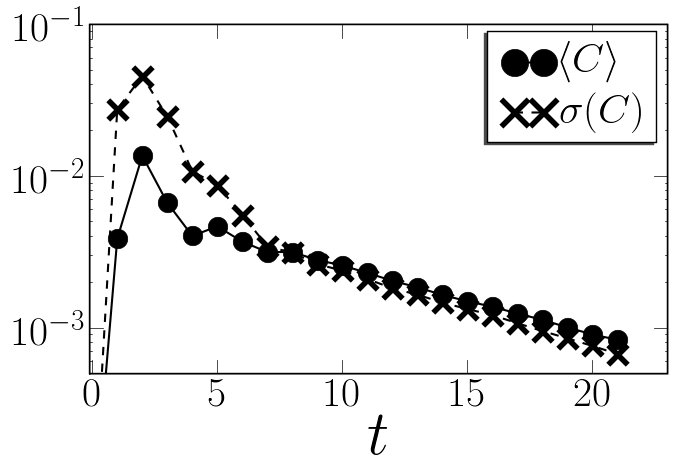}}
}
\end{minipage}
\begin{minipage}{0.38\columnwidth}
\subfigure[]{
\centerline{\includegraphics[width=0.99\textwidth]{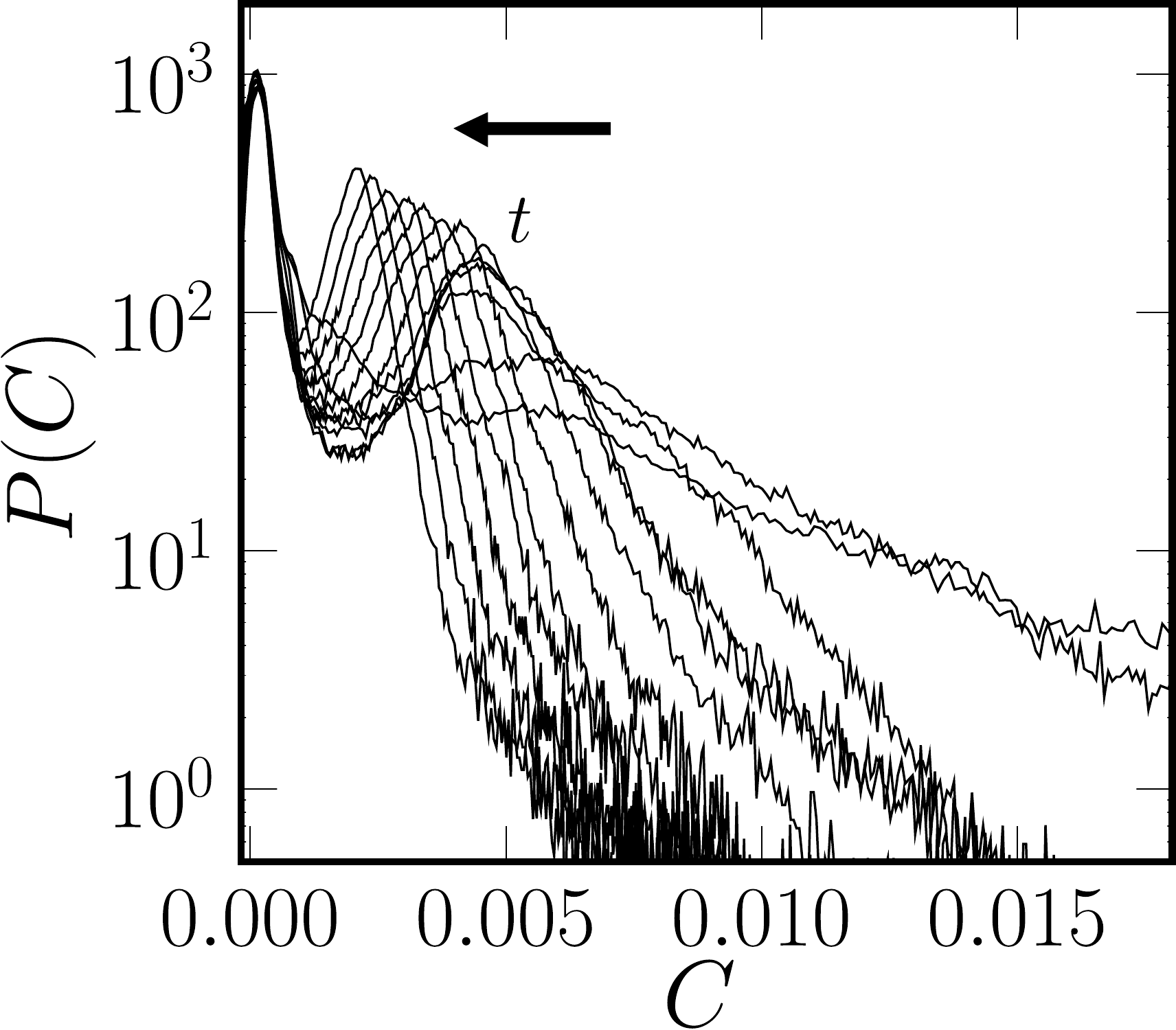}}
}
\end{minipage}
\begin{minipage}{\columnwidth}
\subfigure[]{
\centerline{\includegraphics[width=0.7\textwidth]{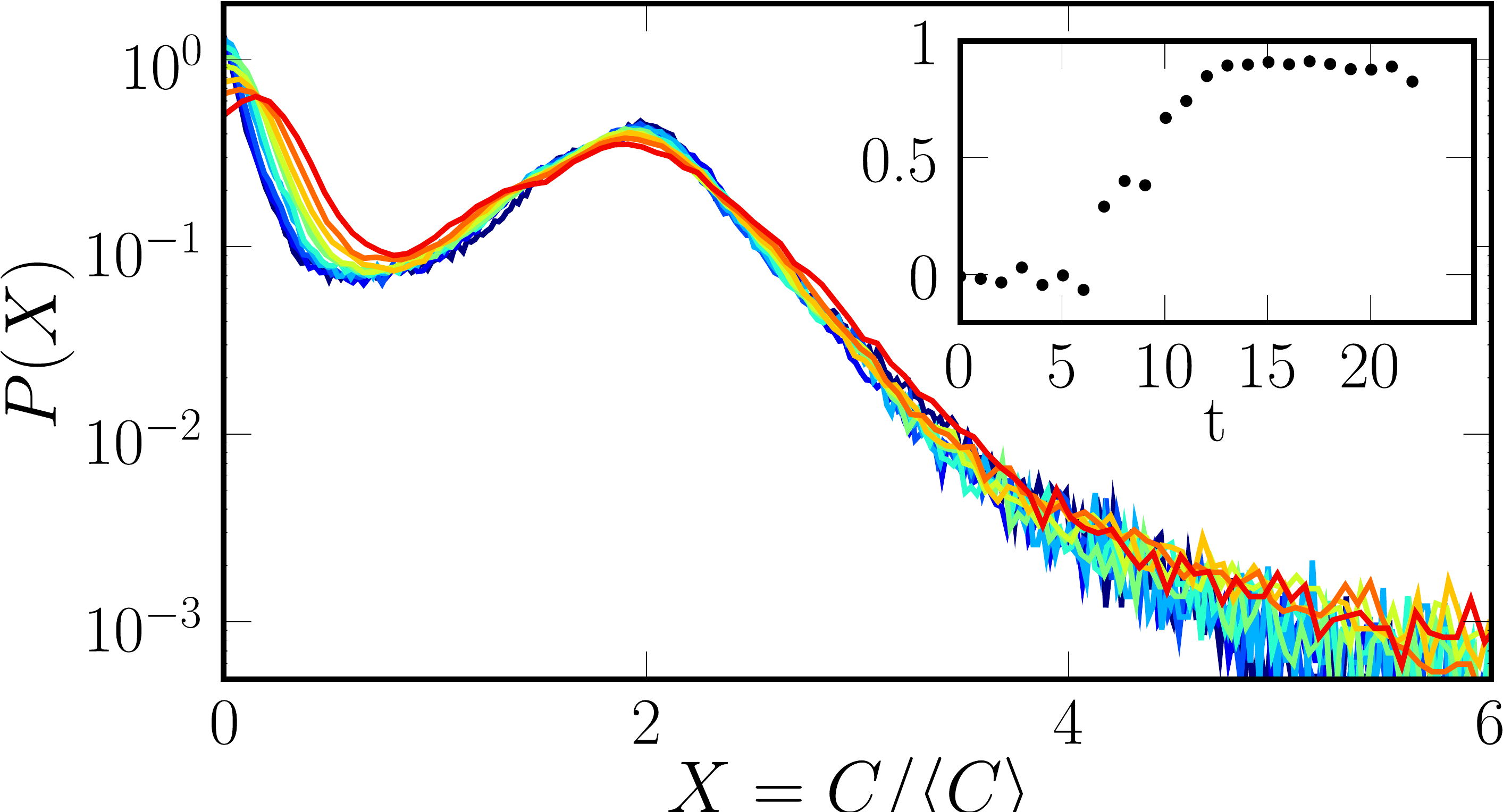}}
}
\end{minipage}
\caption{ (Previous page) \textbf{Butterfly protocol:} (a) Rectangular region of the
mixing pattern where the concentration field is
measured. (b) Evolution of the spatial mean and standard deviation of
the concentration field (in arbitrary units) with the number of stirring
periods $t$, for a non-dimensional mean stirring time $\N=10$. Note the
similar exponential evolution of the first two moments.  (c) Same as
for (b), for $\N=5$. (d) Histograms of the concentration field at
periods 4 to 20
($\N=5$). (e) The histograms for periods 7 to 20 collapse onto a single
curve when the concentration is rescaled by its decaying mean.  Inset in
(d): The time cross-correlation of the concentration pattern
$r(t)=\text{Cov}\!
\left(C(t),C(t+1)\right)/\left(\sigma(t)\sigma(t+1)\right)$ converges to
unity, showing that the spatial pattern repeats over time.
\label{fig:BF}}
\end{figure}

We measure the concentration field of dye in a rectangle inside the
mixing region (Fig.~\ref{fig:BF}(a)). The concentration mean and standard
deviation over the studied domain are plotted for two different stirring
frequencies in Fig.~\ref{fig:BF}(b) and (c). The first thing to note is
that the evolution of the mean is exponential in time, which amounts to
an exponential distribution of residence times inside the mixing region
for dye particles injected in the initial blob. The exponential evolution
of the mean can be interpreted as a feature of chaos: a lobe is advected
away from the mixing region at each period, but chaotic advection folds
the remaining dye over the entire mixing region, so that the same
fraction of the dye is taken away at each period. Residence
time distributions with exponential tails have been observed in several
studies of chaotic advection in open flows~\cite{Jung1993, Sommerer1996},
at least for intermediate times.

Note that the decay rate is controlled to first order by mass
conservation across the mixing region. If all fluid particles were to
undergo a transient chaotic motion and be stretched and folded by the
rods, the decay rate would depend only on the flowrate
and the extent of the mixing region. In the limit of an infinite stirring
frequency, this situation would correspond to a ``perfect
mixer''~\cite{Danckwerts1953, Denbigh1984}, in the language of chemical
engineering. Perfect mixers have the property that a particle leaves the
mixing region with the same probability distribution as if it were picked
at random, and have a purely exponential distribution of residence times.
However, our device cannot be considered as a perfect mixer as some fluid
particles escape rapidly downstream along the channel sides, and
contribute to the short-time peak of the concentration mean in
Fig.~\ref{fig:BF}(b) and (c). The deviation from a purely exponential
distribution of residence times may be considered as a crude measure of
the deviation from perfect mixing.

We also observe on Fig.~\ref{fig:BF}(b) and (c)
that the evolution of the standard deviation is exponential as well,
and \emph{with the same rate} as the mean concentration, a first
symptom of self-similarity.  This is confirmed by the plots of the
concentration histograms taken at different periods
(Fig.~\ref{fig:BF}(d)), which collapse on a single curve when rescaled
by the exponentially-decaying concentration mean
(Fig.~\ref{fig:BF}(e)).  We also observe that the spatial correlation
of patterns taken at successive periods (inset in
Fig.~\ref{fig:BF}(e)) converges rapidly to unity, indicating a
persistent pattern.  This means that after a short transient phase the
evolution of the concentration field is self-similar, and is therefore
described by the exponential decay of its mean:
\begin{equation*} 
\C(\xb, t) = \langle C \rangle(t) \times \Cself(\xb) = \Cinf \,
\exp(-t/\tau) \Cself(\xb).
\end{equation*} 
Such a permanent field is the signature of a \emph{coherent} process (here, a
time-periodic flow), in contrast to e.g. \emph{random} turbulent mixing.

We have thus provided quantitative evidence of a concentration
eigenmode in chaotic advection in open flows.  Note that an eigenmode
scenario was also observed for other stirring frequencies for the
butterfly protocol.  The next section is devoted to an interpretation
of the eigenmode.


\section{An open-flow baker's map model}
\label{sec:BM}

Since Pierrehumbert drew attention to invariant patterns and strange
eigenmodes~\cite{Pierrehumbert1994}, much work has been devoted to their
characterization in closed flows. The mean concentration is conserved in
a closed geometry, so the evolution of the eigenmode corresponds to an
exponential decay of fluctuations around the mean by averaging the
concentration values of neighboring filaments~\cite{Villermaux2003,
Thiffeault2004}. It is now widely believed that the decay rate arises
from a complex combination of spatial correlations of stretching, and
cannot be easily predicted~\cite{Antonsen1996, Wonhas2002, Fereday2002,
Thiffeault2003, Fereday2004, Thiffeault2004b}. In a recent
paper~\cite{Gouillart2008a}, we have shown that the unstable manifolds of
least-unstable periodic orbits are the backbone where the strange
eigenmode takes significant values. In the vicinity of these orbits,
particles repeatedly experience a lower stretching than average, and
concentration fluctuations can survive longer.

\begin{figure}
\subfigure[]{ 
\centerline{\includegraphics[width=0.75\columnwidth]{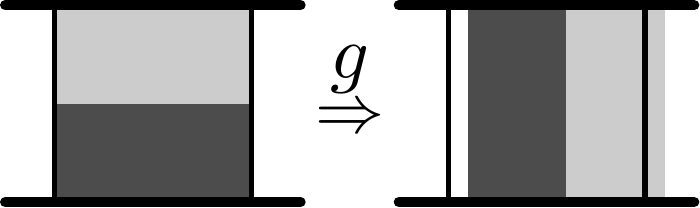}}
}
\subfigure[]{
\centerline{\includegraphics[width=0.9\columnwidth]{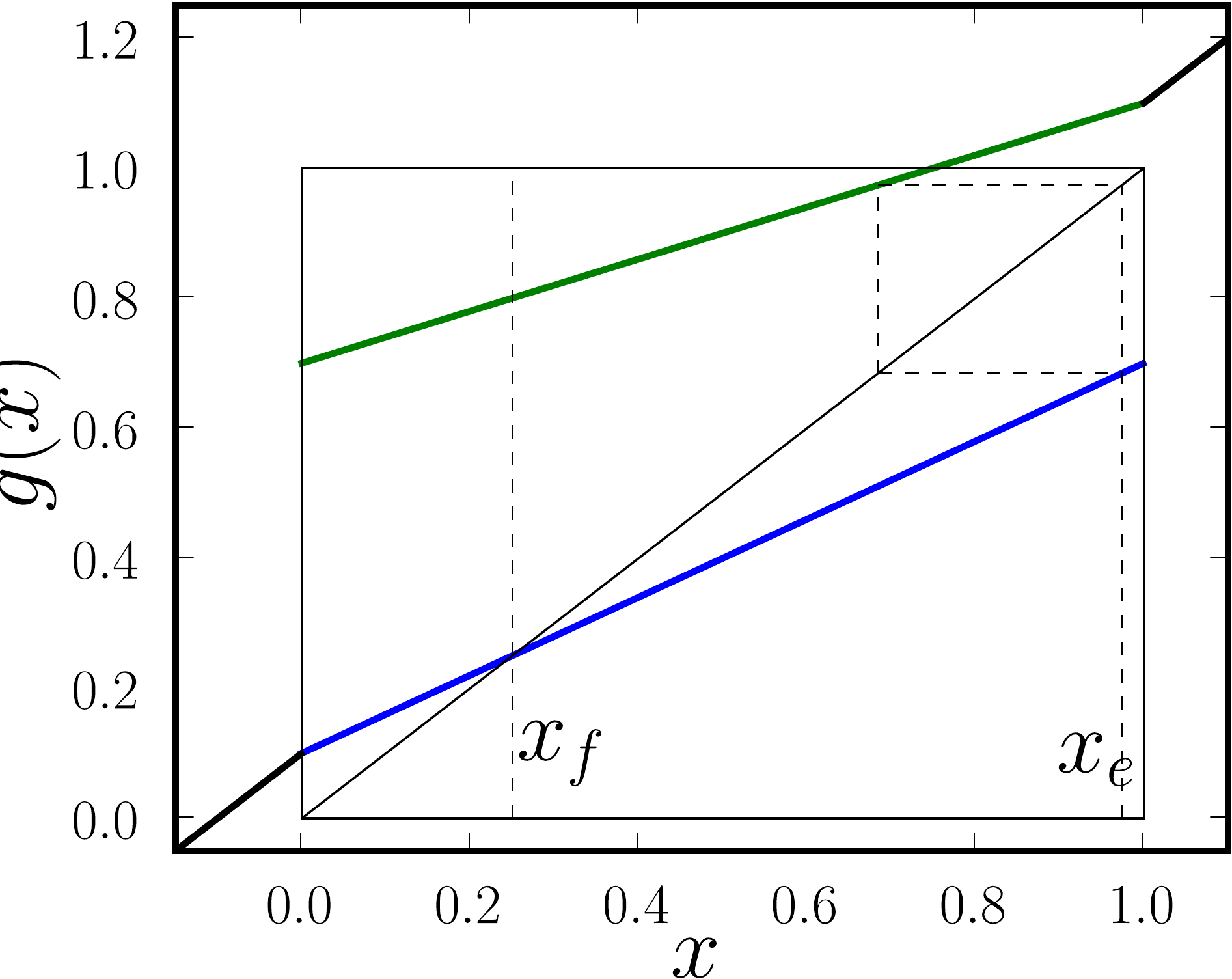}} 
}
\caption{(a) The open-flow baker's map: at each iteration, the unit
  square is cut into two horizontal strips, which are then stretched
  and restacked together in a square translated by $\U$. Outside the
  mixing region, fluid is simply translated by $\U$. (b) Plot of the
  one-dimensional open-flow baker's map $\BM$ for $\gamma=0.6$ and $\U
  = 0.1$. \label{fig:open_map}}
\end{figure}

\begin{figure}
\subfigure[]{
\centerline{\includegraphics[width=0.99\columnwidth]{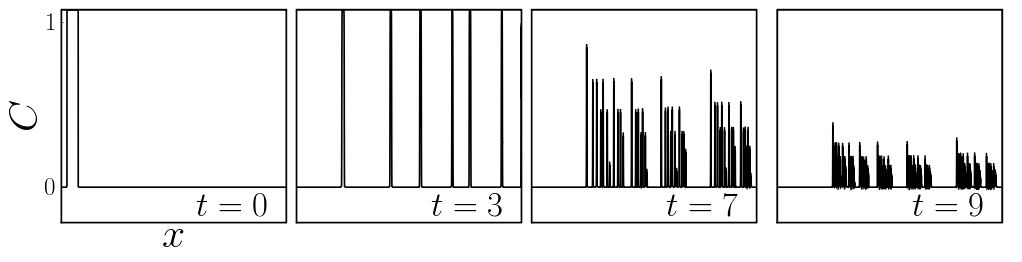}}
}
\subfigure[]{
\centerline{\includegraphics[width=0.9\columnwidth]{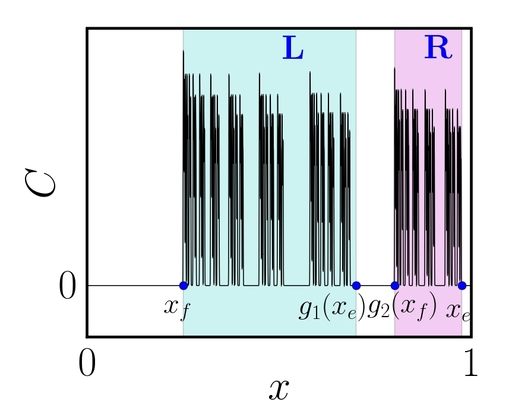}}
}
\caption{(a) Typical evolution of the concentration profile during the
mixing of a blob of dye (left) by the open-flow baker's map. Once
iterates of the initial strip reach the Batchelor scale~$\wb$, the pattern
repeats over time, with an exponentially decaying amplitude. (b) The
support of the concentration profile may be split into two blocks
$\Le$ and $\Ri$, delineated by the iterates of the periodic points that
mark the borders of the chaotic saddle. The amount of dye inside the two
blocks obeys simple symbolic dynamics (Eq.~\eqref{eq:symbol}),
which allows to obtain easily the decay rate of the profile.
\label{fig:profile}}
\end{figure}

In this section, we account for the appearance of
the concentration eigenmode observed in our experiments in open flows.
We will rely on a simplified model for an open flow with
chaotic advection. Our model is a variant of one of the most
studied models of chaotic mixing, the inhomogeneous area-preserving baker's
map~\cite{Farmer1983,Ott1989,Antonsen1991,Fereday2002}.

The open-flow baker's map is defined on a channel-like domain, that is an
infinite strip (Fig.~\ref{fig:open_map}) composed of a square central
mixing region, and corresponding upstream and downstream regions. At each
iteration, the action of the map consists of two parts: (i) a
translation of the strip by a distance $\U$,
mimicking the global advection in an open flow; (ii) a `traditional'
baker's map, that is a division into two strips, which are then
compressed and re-stacked as in Fig.~\ref{fig:open_map}, preserving
area.  The map has the property of mapping a distribution invariant
along the direction~$y$ transverse to the channel into another such
distribution.

We take our initial ``blob of dye'' to be a uniform strip in
the~$y$-direction, i.e. a strip that extends over the whole channel
width.  Inside the mixing region, the baker's map stretches and folds
this strip to create more strips, leaving the concentration
independent of~$y$~\cite{Fereday2002,Fereday2004}.  We can thus focus
on one-dimensional distributions that depend only on the~$x$
coordinate: they represent a `cut' across a striated pattern such as
in Fig.~\ref{fig:BFBS}(c) or (d).  Hence, in the following calculation
we limit ourselves to a one-dimensional version of the baker's map.

Inside the unit interval, the baker's map~$\BM$ reads
\begin{subequations}
\begin{equation}
\BM: x \mapsto \BM_1(x) \cup \BM_2(x),\qquad 0\le x\le 1,
\label{eq:map_baker_a}
\end{equation}
where 
\begin{equation}
\BM_1(x) = \U + \gamma x\,;\qquad \BM_2(x) = \U + \gamma + (1-\gamma)x
\end{equation}
\label{eq:map_baker}%
\end{subequations}%
and the union ($\cup$) symbol in~\eqref{eq:map_baker_a} means
that~$\BM$ is one-to-two: every point~$x$ has two images given
by~$\BM_1(x)$ and~$\BM_2(x)$. The parameter~$\gamma$
satisfies~$0<\gamma<1$ and controls the homogeneity of stretching,
with~$\gamma=1/2$ being the perfectly homogeneous case. We also define
the mean stretching rate or Lyapunov exponent,
\begin{equation*}
  \log\Gamma = \gamma\log\gamma + (1-\gamma)\log(1-\gamma).
\end{equation*}
In the map $\U$ represents the flow rate, where the unit of time is the
period of the map, and the unit of length the width of the
channel. The baker's map $\BM$ is represented in
Fig.~\ref{fig:open_map}(b) for the reference values we will use
throughout the paper, $\gamma=0.6$ and $\U = 0.1$.

Under the action of the baker's map, the concentration profile evolves
as
\begin{equation}
C(x,t+1)=C\bigl(\BM^{-1}(x),t\bigl).
\label{eq:Cevolve}
\end{equation}
The map $\BM$ therefore transforms the concentration profile $C(x,t)$
at time $t$ into two images compressed by a factor $\gamma$ and
$1-\gamma$, respectively.  In contrast to the classical baker's map, a
fraction of the strip is mapped outside of the unit interval at each
iteration, so that the total amount of ``dye'' in the central interval
decreases. Note that ``leaking'' baker's maps relying on similar
principles have already been used in~\cite{Tel1996, Neufeld1998,
  Schneider2002}; however, our map follows more closely the geometry
of the channel flow.  Diffusion is mimicked by letting the
concentration evolve diffusively during a unit time
interval~\cite{Pierrehumbert2000,Fereday2002}.  During that interval,
$C$ evolves according to the heat equation with diffusivity
$\kappa$. 

First, we compute numerically the evolution of an initial blob under
the action of $\BM$.  The initial condition (leftmost frame in
Fig.~\ref{fig:profile}(a)) is a strip of constant concentration
$\C=1$, located in the interval $[0, \U]$.
A few iterates ($t=0,3,7,9$) of the concentration
profile are shown in Fig.~\ref{fig:profile}(a) for the parameters
$\gamma = 0.6$, $\U=0.1$. The initial blob is repeatedly stretched and
folded into many thinner strips (second to fourth frames in
Fig.~\ref{fig:profile}(a)). Some of the strips are mapped downstream,
so that the number of remaining iterates is less than~$2^t$ at
time~$t$ --- or equal to~$2^t$ for a vanishing flowrate~$\U=0$.

As soon as strips are compressed to the Batchelor scale~$\wb$,
diffusion starts smoothing the strips, so that the profiles of
neighboring strips are merged together. From this moment on, we
observe the recurrence of the same concentration pattern, but with
lower amplitude at each iteration (third and fourth plots in
Fig.~\ref{fig:profile}(a)). This eigenmode behavior is confirmed by
the same statistical analysis as in experiments: as shown in
Fig.~\ref{fig:stats_BM}(a), we measure an exponential decay of the
concentration mean inside the mixing region, and a similar decay of
higher moments.  Moreover, concentration histograms collapse onto a
single curve when the concentration is rescaled by its mean value
(Fig.~\ref{fig:stats_BM}(b)). The map therefore reproduces well the
salient features observed in the physical flow.

\section{Symbolic dynamics of the open-flow baker's map
\label{sec:symbol_dyn}}

We now exploit the simplicity of the map to gain more insight into the
structure of the asymptotic concentration pattern, and to describe its
evolution with time. Periodic points of the chaotic saddle play a
special role in the concentration profile, as they form the backbone
of the profile. Some iterates of dye particles converge rapidly
towards periodic orbits and remain close to them. The support of the
asymptotic profile is therefore the set of periodic points broadened
by diffusion to the minimal scale $\wb$. In other words, the mixing
profile is composed of a succession of iterates of the interval
$[0,\U]$ that have entered the mixing region at different times: apart
from a zero-measure set of periodic orbits --- the chaotic saddle ---
each point of the interval can be mapped backward in time to a point
in the upstream region. Iterates of $[0,\U]$ wider than $\wb$ do not
overlap with other iterates and appear as gaps in the pattern (no
periodic points are present in these intervals), while iterates that
have been compressed down to $\wb$ overlap with images of the initial
strip and belong to regions where the concentration is positive ---
the fractal chaotic saddle~\cite{Tel2000}.  

The central mixing region encloses the support of the chaotic saddle, and
extends from~$x=\xf$ to~$x=\xe$, $\xf<\xe$ (see Fig.~\ref{fig:profile}
(b)).  Here, $\xf\equiv \U/(1-\gamma) = 0.25$ is the unique fixed point
of the map, and~$\xe$ is the rightmost periodic point in the chaotic
saddle.  For our parameter values,~$\xe$ belongs to the unique period-2
orbit of the map, $\xe = (\U(\gamma-2)+\gamma)/(1+\gamma(\gamma-1))
\simeq 0.974$.

\begin{figure*}
\begin{minipage}{0.48\columnwidth}
\centerline{\includegraphics[width=0.98\columnwidth]{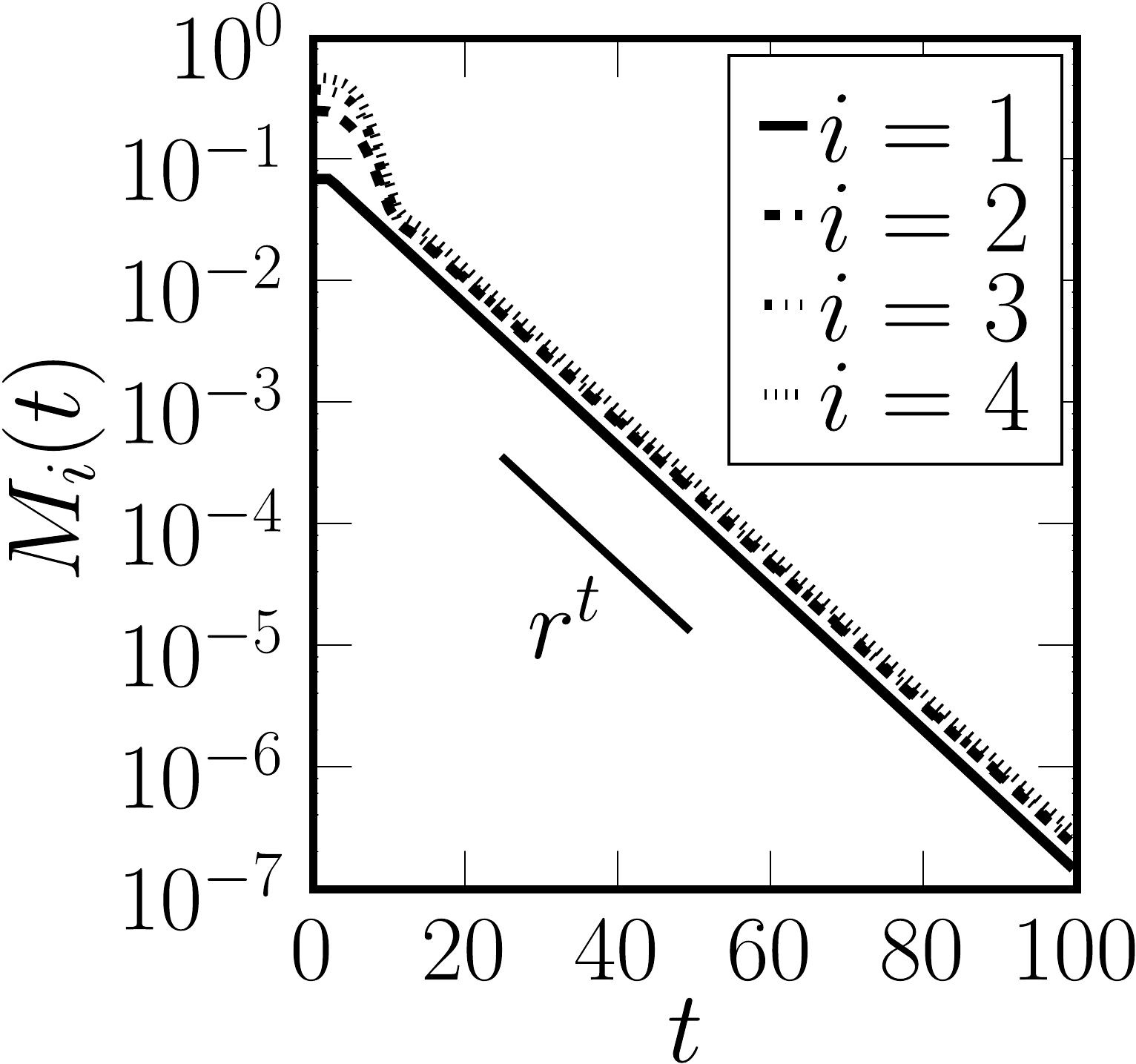}}
\end{minipage}
\begin{minipage}{0.48\columnwidth}
\centerline{\includegraphics[width=0.98\columnwidth]{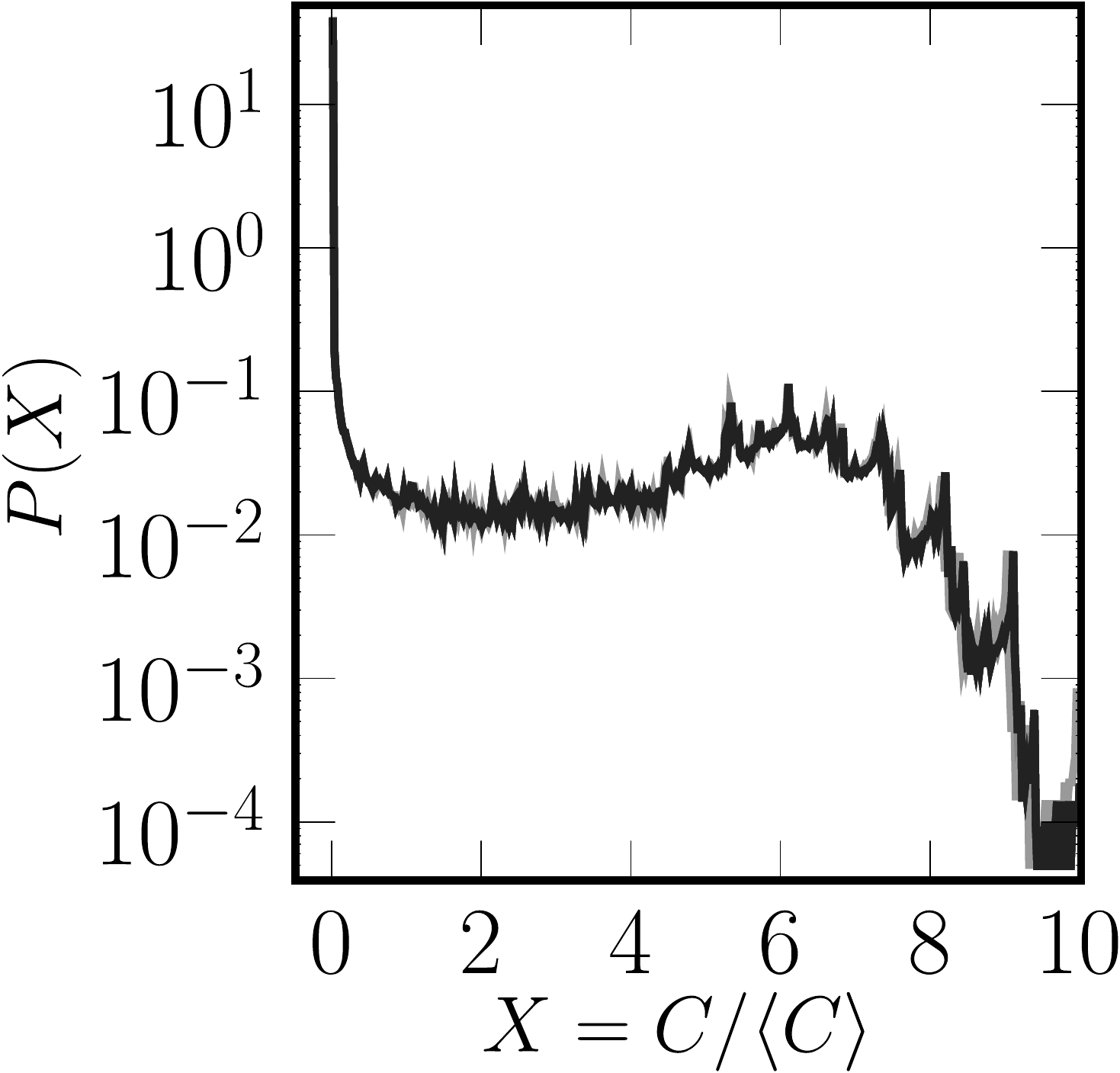}}
\end{minipage}
\caption{\textbf{Open-flow baker's map:} (a) Evolution of the first
  moments of the concentration field $M_i(t)=\bigl|\int_0^1
    (C(x)-\langle C \rangle)^i\d x\bigr|^{1/i}$ under the action of
  the open-flow baker's map. The mean concentration decays
  exponentially almost from the injection of the dye, while the other
  moments adopt a similar exponential evolution after a short
  delay. (b) Concentration histograms at periods
  $15,\,25,\,\cdots,\,95,\,$ collapse very well on a single
  curve.\label{fig:stats_BM}}
\end{figure*}

We now look for a partition of the unit interval into disjoint regions
that are mapped onto each other by the map --- that is, we search for
a \emph{Markov partition} of the map~\cite{Adler1998, Katok1995},
which describes succinctly the dynamics of the map.  The partition is
particularly simple in our example: using the map
expression~\eqref{eq:map_baker}, one easily checks that defining
respectively the blocks $\Le$ and $\Ri$ as the intervals $[\xf,
\BM_1(\xe)]$ and $[\BM_2(\xf), \xe]$ divides the concentration pattern in
two disjoint blocks. Moreover, $\BM_1$ compresses both $\Le$ and $\Ri$
inside $\Le$, whereas $\BM_2$ moves a compressed image of $\Le$ to the
location of $\Ri$, and advects $\Ri$ downstream. Defining as $\Le_t$ and
$\Ri_t$ the amount of dye remaining in the two blocks at time $t$, we
obtain the symbolic dynamics~\cite{Beigie1991b,Tel1996}
\begin{subequations}
\begin{align}
\Le_{t+1} &= \gamma\left[\Le_t + \Ri_t \right], \\
\Ri_{t+1} &= (1-\gamma) \Le_t. 
\end{align}
\label{eq:symbol}%
\end{subequations}%
The symbolic dynamics~\eqref{eq:symbol} explains the exponential decay
of the concentration mean: after a short transient, the total amount
of dye $\Le_t+\Ri_t$ 
decays exponentially at a rate $\log\R$, where
\begin{equation}
\R = \tfrac12\bigl(\gamma + \sqrt{4\gamma -3\gamma^2}\bigr) \simeq 0.874
\end{equation}
is the largest eigenvalue of the linear system~\eqref{eq:symbol}.
This is indeed the rate of decay measured in simulations
(Fig.~\ref{fig:stats_BM}(a)).

Symbolic dynamics therefore allows us to determine the exact form of the
global asymptotic evolution of the concentration profile in the map. Note
that the value of $\R$ depends only on the specific form of symbolic
dynamics~\eqref{eq:symbol}. In particular, $\R$ does not depend on $\U$
inside the interval where the symbolic dynamics~\eqref{eq:symbol} is
valid. For example, for a fixed value of $\gamma$, the symbolic
dynamics~\eqref{eq:symbol} accounts for the map $\BM$ in the interval
\begin{equation} \frac{(1-\gamma)^2}{3-2\gamma} < \U <
\frac{(1-\gamma)^2}{2-\gamma}. \label{eq:condU} \end{equation} We have
checked numerically that the dependence of~$\R$ on~$\U$ follows a
devil's staircase, that is a function constant almost everywhere except
at values where the form of the symbolic dynamics change. This happens
when the period of the rightmost periodic orbit $\xe$ changes, as the
rightmost iterate of the interval $[0,\U]$ gradually slides to the right
as $\U$ is increased, until it leaves completely the unit interval and it
is replaced by another iterate with different symbolic dynamics. There is
no obvious correlation in the successive periods of the rightmost
orbits as $\U$ is increased. Symbolic dynamics where the rightmost
orbit has a short period are valid for a wider range of $\U$ values
than for longer periods.

The global exponential decay of the total concentration value also
imposes the same exponential decay everywhere in the profile, hence
the onset of a strange eigenmode. This can be explained as
follows. Choose a box centered on $x$ of size given by the local scale
of variation of the concentration profile $\wb$, and suppose we
measure a positive value of the concentration $\C(x)$ in this
box. After a typical number of periods $n=\log \Gamma / \log \wb$
(which is the number of iterations needed to compress the whole unit
interval to the Batchelor scale), the concentration value is the
average of many images of the initial profile. Neglecting round-off
and truncation errors, we may consider that $C$ measured at time $t$
represents the total amount of dye inside the whole unit interval at
time $t - n$, multiplied by the compression factor $\Gamma^{n}$, that
is a quantity
\[
C(x,t)\simeq \Gamma^{n}(\Le_{t - n} + \Ri_{t - n}).
\]
In the next period $t+1$, the value of the concentration measured at $x$
is now
\[
C(x,t+1)\simeq \Gamma^{n}(\Le_{t+1 - n} + \Ri_{t+1 - n}),
\]
that is
\begin{equation}
C(x,t+1) = \R C(x,t)
\label{eq:Ceigen}
\end{equation}
as we have shown that $\Le_t$ and $\Ri_t$ have a geometric progression.
Eq.~\eqref{eq:Ceigen} is exactly the equation of an eigenmode (or
invariant distribution) of the advection operator (the map). While not
rigorous, the above reasoning shows that a concentration eigenmode
appears because of the coherent cascading of large blocks towards smaller
scales resulting from the repeated compression by the same map, whereas
the decay of the total concentration in the larger blocks is explained
by considering symbolic dynamics that describe the map. 

We have chosen a set of parameters where the Markov partition of the
map, and consequent symbolics dynamics, are very simple. Nevertheless,
the same line of reasoning can be followed for more complicated
partitions to explain how a concentration eigenmodes sets in. In the
experiments, we know from the study of surface
diffeomorphisms~\cite{Boyland2000} that it is possible to define a
Markov partition with a finite number of blocks for the real flow as
well --- in the absence of elliptical islands.  Hence, we can infer
the onset of an eigenmode from the existence of such a
partition. However, it is very difficult to find the location of periodic
orbits and the Markov partition
from experimental pictures. In contrast, for the map, the partition can
be determined by partitioning the unit interval using the iterates of
periodic points~\cite{Solari2005}.


\section{Deviations from the strange eigenmode 
}
\label{sec:BS}

In sections~\ref{sec:decay_BF}  and~\ref{sec:BM}, we showed the appearance
of a concentration eigenmode for experiments conducted with the butterfly
protocol, where the chaotic saddle is located far from the channel walls,
and we used an open-flow baker's map to provide a simple physical
picture of this evolution.  We now report on experiments
where the concentration pattern does not converge to an eigenmode, which
we attribute to the non-hyperbolic character of the chaotic saddle.

\begin{figure}
\begin{minipage}{0.48\columnwidth}
\subfigure[]{
\centerline{\includegraphics[width=\textwidth]{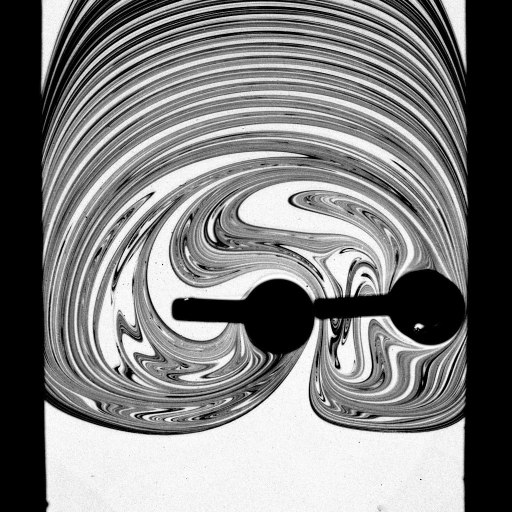}}
}
\end{minipage}
\begin{minipage}{0.48\columnwidth}
\subfigure[]{
\centerline{\includegraphics[width=\textwidth]{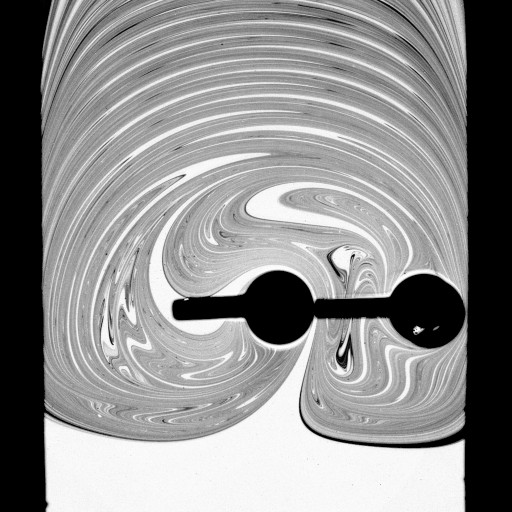}}
}
\end{minipage}
\begin{minipage}{0.48\columnwidth}
\subfigure[]{
\centerline{\includegraphics[width=\textwidth]{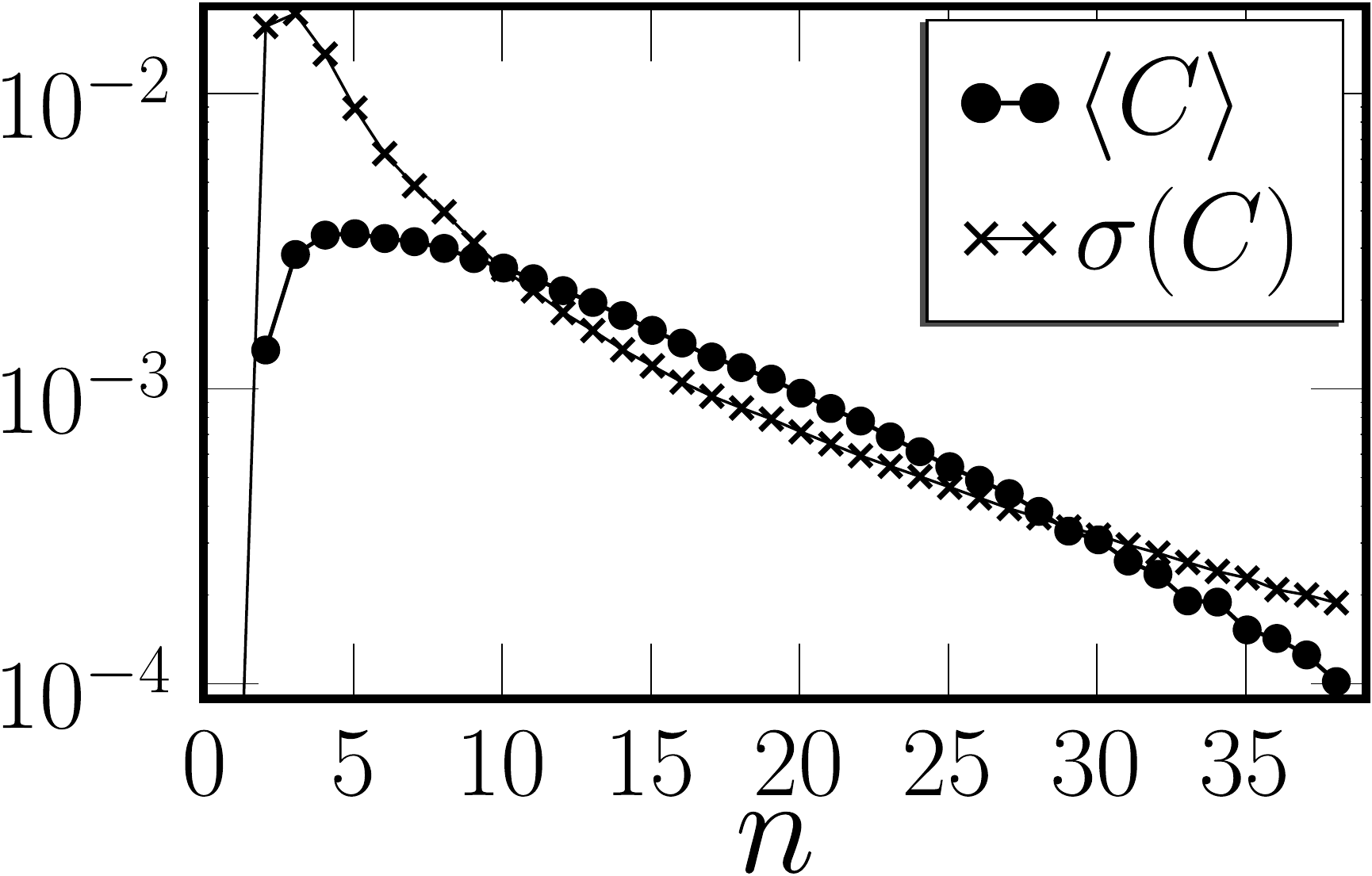}}
}
\end{minipage}
\begin{minipage}{0.48\columnwidth}
\subfigure[]{
\centerline{\includegraphics[width=\textwidth]{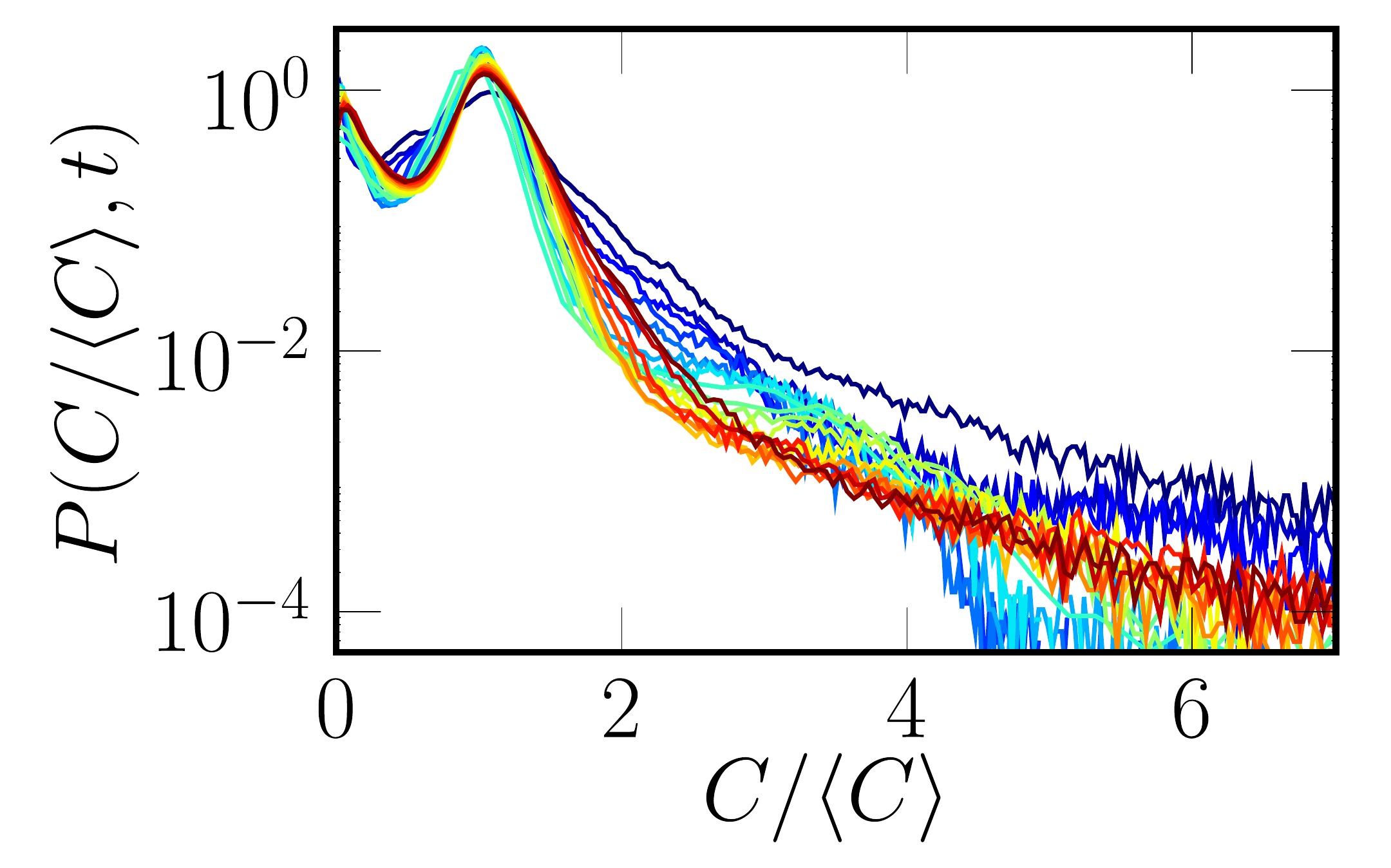}}
}
\end{minipage}
\caption{[Color online] \textbf{Breaststroke protocol:} Concentration
  pattern (a) 12 and (b) 15 periods after the initial blob enters the
mixing region. Beyond the obvious loss of dye, the shape of the two
patterns differs slightly.  The dye approaches the unstable manifold of
the two separation points and is increasingly concentrated in the
vicinity of the manifold. (c) Time-evolution of the mean and standard
deviation of the concentration field (in arbitrary units).  In contrast
to the butterfly protocol, the moments evolve differently and
non-exponentially. (d) Concentration histograms, plotted here for periods
14 to 34, cannot be rescaled on a single curve as for the butterfly
protocol.  The concentration field therefore does not converge to an
eigenmode for the duration of an experiment. \label{fig:BS} }
\end{figure}

We first present results for the evolution of the concentration field
in breaststroke experiments (see Figs.~\ref{fig:BFBS}(b) and (d) for a
description of the rod motion). As for the butterfly protocol, we
study the evolution of the first moments and histograms of the
concentration, respectively plotted in Figs.~\ref{fig:BS}(c) and (d)
for a typical experiment. The evolution of the concentration mean
plotted in Fig.~\ref{fig:BS}(c) is close to exponential,
though somewhat faster.  This means that dye increasingly concentrates
in a zone where it is rapidly advected downstream by the flow.
However, the standard deviation, also plotted in Fig.~\ref{fig:BS}(c), does not follow the evolution of
the mean: it is slower than an exponential law.  In addition,
concentration histograms taken at different periods do not collapse
onto a single curve: Fig.~\ref{fig:BS}(d) reveals that the culprit is
the high-concentration tail of the histogram, whose relative
importance grows with time.  This is the signature of dark filaments
reinjected along the unstable manifolds of separation points (see
Sec.~\ref{sec:mix_patterns} for a discussion of separation points).

Hence, in the breaststroke protocol the concentration is not an
eigenmode, since it is neither repeating nor exponentially-decaying. This
can be related as follows to the concentration patterns discussed in
Sec.~\ref{sec:mix_patterns}.  We have already observed that mixing
patterns for the breaststroke protocol differ significantly from those
for the butterfly protocol. The patterns extend over the entire channel
width, and the boundary between the upstream region and the mixing
pattern is fixed by two \emph{separation points} and their unstable
manifolds. Following earlier work by Chertkov et al.~\cite{Chertkov2003a,
Lebedev2004}, we have described in recent work \cite{Gouillart2007,
Gouillart2008a} how Lagrangian particles come closer to, or escape from,
separation points obeying slow algebraic dynamics. As a consequence of
the no-slip conditions, separation points on the walls are parabolic
points that are only marginally unstable~\cite{Jana1994} and therefore
slow down fluid in their vicinity~\cite{Gouillart2007, Gouillart2008a}.
Particles approach the wall according to slow algebraic dynamics
$d(t)\sim 1/t$ for asymptotic times. A hint of such slow dynamics is
evident in successive pictures of the mixing pattern, which grows slowly
towards the unstable manifold of the separation points
(Figs.~\ref{fig:BS}(a) and (b)). In contrast, for butterfly protocols the
mixing pattern rapidly adopts a constant shape.

Dye filaments from the initial blob are therefore trapped for long times
in the vicinity of separation points, where they are hardly stretched and
mixed. Unmixed fluid is slowly reinjected along the unstable manifold of
separation points, forming dark filaments clearly visible in
Fig.~\ref{fig:BS}. As time goes on, dye is almost exhausted inside the
bulk, whereas there remains a pool of poorly mixed fluid near the
separation points that slowly leaks into the bulk. This contrast between
bulk and reinjected filaments intensifies with time
(Figs.~\ref{fig:BS}(a) and (b)), in contradiction with the hypothesis of
a decaying eigenmode. The discrepancy with a concentration eigenmode may
therefore be traced back to the differences in the stretching rate
near the walls, where stretching vanishes, and the bulk. 

\begin{figure}
\subfigure[]{
\centerline{\includegraphics[width=0.6\columnwidth]{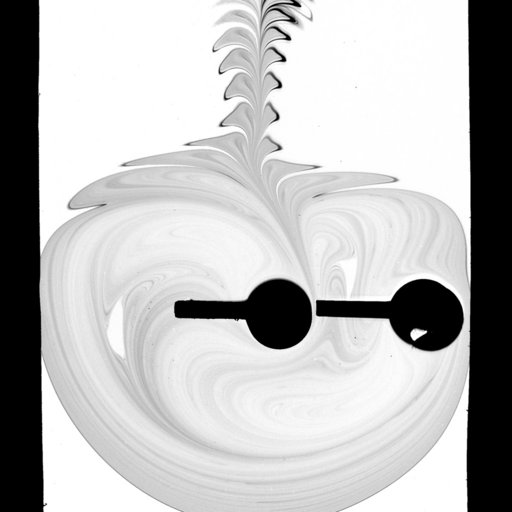}}
}
\subfigure[]{
\centerline{\includegraphics[width=0.7\columnwidth]{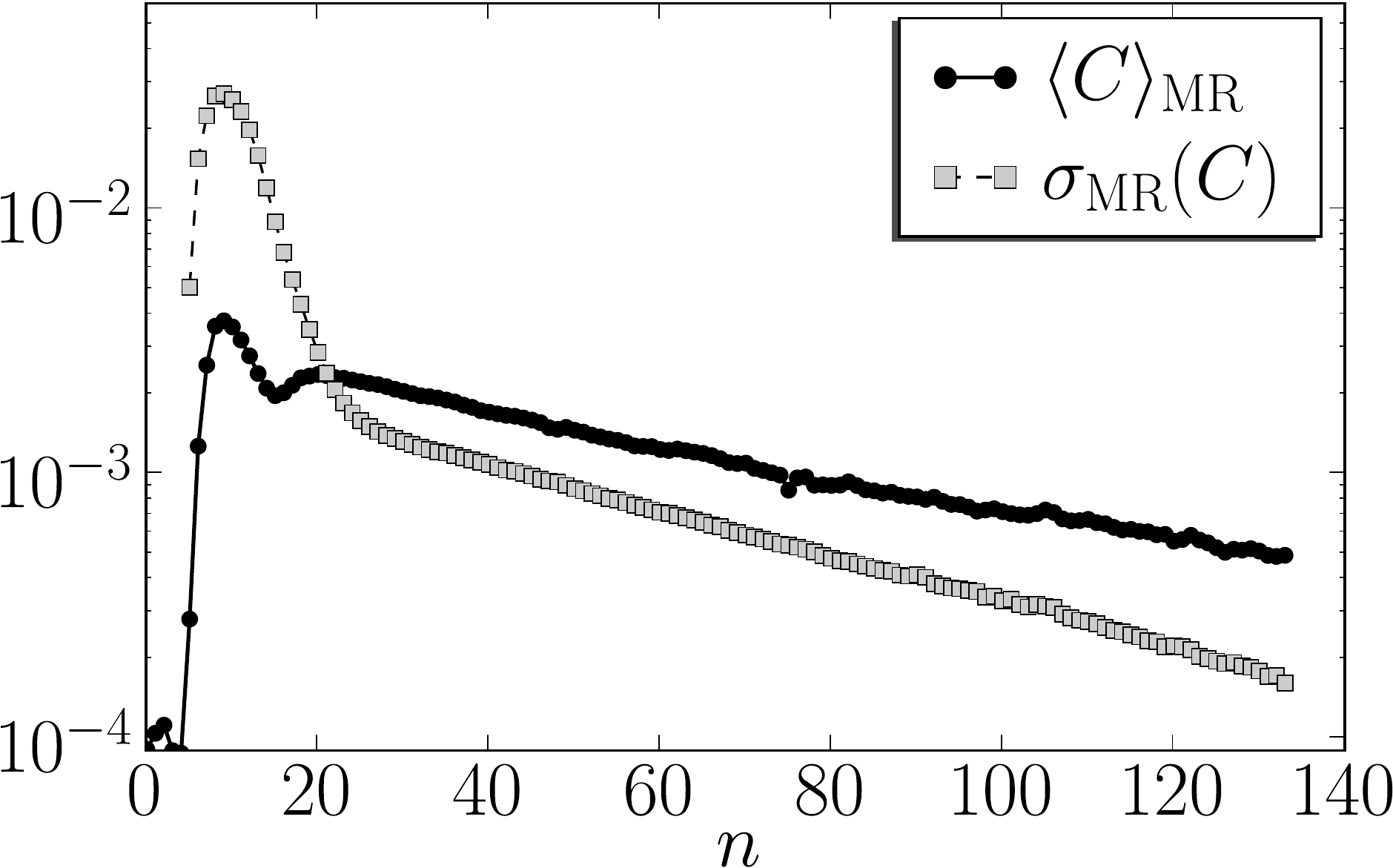}}
}
\caption{(a) Mixing pattern for a high stirring speed relative to the
mean velocity of the flow.  Two elliptical islands are visible close to
the center of the circular paths of the rods. ``Branches'' of higher
contrast emanate from the islands as a consequence of their
stickyness. (b) The time-evolution of the mean and
standard deviation of the concentration field are exponential, but with
different exponents for the two moments. This departure from an eigenmode
behavior may be attributed to the presence of the islands, which retain
dye in their vicinity for long times.\label{fig:BF8}}
\end{figure}

Structures of the phase space other than parabolic separation points may
cause departure from an eigenmode behavior of the concentration pattern.
It is well known that elliptical islands can store poorly-mixed fluid in
the vicinity of their boundary, a phenomenon known as \emph{stickyness}
of islands~\cite{Zaslavsky2002, Motter2005}.  This stickiness is
associated with exceptionally-low stretching values near the island. For
the butterfly protocol, increasing the stirring frequency causes two
small islands to appear inside the circles trace out by the rods
(Fig.~\ref{fig:BF8}).  These stable islands are destroyed by the main
flow at lower stirring frequency.  We indeed observe that dye accumulates
around the islands with time, whereas it is depleted in the chaotic bulk.
Dye is however slowly fed from the islands into the bulk by filaments
escaping from the islands, visible in Fig.~\ref{fig:BF8}.  This evolution
of the pattern is incompatible with the onset of a strange eigenmode: the
evolutions of the mean and standard deviation of the concentration are
not similar.

\section{Conclusions}
\label{sec:concl}

We have studied the mixing of a dye tracer by chaotic advection in an
open-flow geometry, where a net free-surface flow crosses a limited
stirring region. Our device was designed to be relevant for industrial
applications, but our foremost objective was to classify mixing
protocols according to generic mixing scenarios. For a first class of
mixers characterized by the absence of regular islands in the mixing
pattern or separation points on the walls, we have observed permanent
concentration patterns that decay exponentially with time, and uniformly
in space. A quantitative analysis of the concentration field confirmed
that the same pattern repeats over time with decaying intensity. The
onset of a permanent pattern may be attributed to the convergence of the
concentration field to an eigenmode of the advection-diffusion operator,
as it is now commonly accepted for closed flows~\cite{Pierrehumbert1994}.
We have provided here a more physical picture of the eigenmode by
introducing an open-flow variant of the baker's map. Symbolic dynamics of
the map shed further light on the global decay of the total concentration
in the mixing region. The repeated and coherent compression of the whole
concentration profile down to smaller scales then explains the global
decay at the same rate everywhere in the profile, hence the onset of a
permanent pattern.

However, we observed a departure from the eigenmode in the case where
stretching is anomalously slow in some regions of the phase space,
that is when either parabolic points on no-slip walls or elliptic
islands are present in the mixing region.  The former case is related
to a recent study~\cite{Gouillart2007, Gouillart2008a} where the
influence of separation points on the dynamics of mixing in closed
flows was investigated.  In the case of a no-slip boundary condition
on fixed walls, separation points promote slow reinjection of fluid
that has been poorly stirred in the vicinity of the walls.  This leads
to a net slowdown of mixing dynamics, yielding algebraic decay instead
of the exponential decay expected for a purely hyperbolic system.

Still, many important questions remain, in particular regarding
mixing dynamics in the breaststroke case. While a previous
study~\cite{Jung1993} reported a power-law shape of the residence-time
distribution in a Von K\'arm\'an alley in the wake of a fixed cylinder, we
have always observed a decay of the mean concentration faster than a
power law. Furthermore, we found the shape of the time-evolution of the
mean concentration to depend on whether the blob of dye had been injected
in the center of the channel or close to a side wall. Further
investigation and modeling of this class of mixers is therefore needed. 

A crucial question in chemical engineering concerns the quantitative
evaluation of the mixing properties of different devices. We have
studied here the decay of the first moments of the concentration
field, which provide information about the time during which a
``contaminant'' remains in the mixing region with a significant
concentration. Other situations of interest for industrial processes
include a continuous injection of inhomogeneous fluid (say a
succession of blobs of dye), where one expects the output to be
``homogeneous enough'' on a large range of scales. Since the work of
Danckwerts~\cite{Danckwerts1953}, a measure of mixing commonly used is
the \emph{intensity of segregation}, which measures the ability of a
mixer to reduce fluctuations (as evaluated by the variance of the
concentration field) between the input and the output. However, this
measure depends strongly on the spatial distribution of the
inhomogeneity.  In contrast, for the butterfly case, the strange
eigenmode characterizes the \emph{mixer itself}, independent of
initial condition. It may therefore be tempting to relate the
characteristics of the strange eigenmode, i.e. its probability
distribution function or its power spectrum, to the mixer's capacity
for reducing the concentration fluctuations. Future work will deal
with the characterization of the early transient pattern, which
consists of the worst-mixed fluid elements, and address the
short-residence-time regime and a subsequent assessment of mixing
quality.

\section*{Acknowledgements}

We gratefully acknowledge enlightening discussions with Franck
Pigeonneau about the form of the channel flow. The authors also thank
C\'ecile Gasquet and Vincent Padilla for technical assistance, as well
as Fran\c{c}ois Daviaud and Philippe Petitjeans for fruitful
discussions.

\bibliographystyle{pfa}
\bibliography{journals_abbrev,mixing}

\end{document}